\RequirePackage{fix-cm}
\documentclass[smallcondensed]{svjour3}     % onecolumn (ditto)
\smartqed  % flush right qed marks, e.g. at end of proof
\usepackage{graphicx}
\usepackage{natbib}
\usepackage{url}
\usepackage{longtable}
\usepackage{amssymb}
\renewcommand{\figurename}{Figure}

\journalname{Solar Physics}
\begin{document}

\title{Ionospheric Disturbances and their Impact on IPS Using MEXART 
Observations%\thanks{Grants or other notes
}


\author{M.\,Rodr\'\i guez-Mart\'\i nez         \and
        H.\,R.\,P\'erez-Enr\'\i quez \and
	A.\,Carrillo-Vargas \and
	R.\,L\'opez-Montes  \and
	E.\,A.\,Araujo-Pradere  \and
	G.\,A.\,Casillas-P\'erez  \and
	J.\,A.\,L.\,Cruz-Abeyro %etc.
}

\authorrunning{M.\,Rodr\'\i guez-Mart\'\i nez {\it et al.}} % if too long for running head

\institute{M.\,Rodr\'\i guez-Mart\'\i nez \at
              Universidad Nacional Aut\'onoma de M\'exico, Escuela Nacional de Estudios Superiores, Unidad Morelia. Antigua Carretera a 
              P\'atzcuaro No. 8701, Ex-Hacienda de San Jos\'e de la Huerta. C.P. 58190. Morelia Michoac\'an, M\'exico.
              %Tel.: +5255 5623-4104, ext-108\\
              %\email{mrodriguez@enesmorelia.unam.mx}           %  \\
           \and
           H.\,R.\,P\'erez-Enr\'\i quez \at
              Universidad Nacional Aut\'onoma de M\'exico, Centro de Geociencias, Campus Juriquilla. Blvd Juriquilla 3001, Juriquilla Quer\'etaro. 76230, 
M\'exico. 
              %Tel.: +5255 5623-4104, ext-106\\
              %\email{roman@geociencias.unam.mx}
           \and
	   A.\,Carrillo-Vargas \at
	      Universidad Nacional Aut\'onoma de M\'exico, Instituto de 
Geof\'\i sica, Unidad Michoac\'an. Antigua Carretera a 
              P\'atzcuaro No.8701, Ex-Hacienda de San Jos\'e de la Huerta. C.P. 58190. Morelia Michoac\'an, M\'exico.
              %\email{armando@geofisica.unam.mx}
           \and
	   R.\,L\'opez-Montes  \at
              Universidad Nacional Aut\'onoma de M\'exico, Centro de Geociencias Campus Juriquilla. Blvd Juriquilla 3001, Juriquilla, Quer\'etaro. 76230, 
M\'exico. 
              %\email{rebeca@geociencias.unam.mx}
           \and
           E.\,A.\,Araujo-Pradere \at
              CIRES-University of Colorado. 325 Broadway W/NP9. Boulder, CO, 
USA. 
              %\email{Eduardo.Araujo@noaa.gov}
           \and
           G.\,A.\,Casillas-P\'erez  \at
              Universidad Nacional Aut\'onoma de M\'exico, Instituto de 
              Geof\'\i sica, Unidad de C\'omputo. Ciudad Universitaria, C.P. 
              04510, M\'exico. 
              %\email{gacp@geofisica.unam.mx}
           \and
           J.\,A.\,L.\,Cruz-Abeyro \at
              Universidad Nacional Aut\'onoma de M\'exico, Centro de Geociencias, Campus Juriquilla. Blvd Juriquilla 3001, Juriquilla Quer\'etaro. 76230, 
M\'exico. \\
             % \email{lcabeyro@geociencias.unam.mx}
             \email{mrodriguez@enesmorelia.unam.mx}
}

\date{Received: date / Accepted: date (will be inserted by the editor) \\
DOI: 10.1007/s11207-014-0496-8.}

\maketitle

\begin{abstract}
We study the impact of ionospheric disturbances on the Earth's 
environment caused by the solar events that occurred from 20 April to 31 May 2010, using 
observations from the {\it Mexican Array Radio Telescope} (MEXART). During this period of 
time, several astronomical sources presented fluctuations in their radio signals. 
Wavelet analysis, together with complementary information such as the vertical total electron 
content ({\it vTEC}) and the {\it Dst} index, were used to identify and understand when the interplanetary scintillation (IPS) could be contaminated by ionospheric disturbances (IOND). We 
find that radio signal perturbations were sometimes associated with IOND and/or IPS fluctuations; however, in some cases, it was not possible to clearly identify their origin. Our Fourier and 
wavelet analyses showed that these fluctuations had frequencies in the range $\approx$\,0.01\,Hz -- $\approx$\,1.0\,Hz (periodicities of 100\,s to 1\,s).
\end{abstract}

\section{Introduction}
\label{intro}
 
Large-scale solar-wind disturbances in the interplanetary medium (IPM) can, and frequently 
do, distort radio-wave fronts coming from compact radio sources producing 
the so-called interplanetary scintillation (IPS) \citep{hew64,hew86}. It is known that radio 
sources with an angular diameter smaller than 1 arcsec present  interplanetary scintillation 
at frequencies $>$\,0.1\,Hz  \citep{hew87}; though, depending on the observing wavelength, IPS 
can be present beyond 10\,Hz \citep{mil76,man90,man94,man10}. IPS on a large number 
of compact radio sources has been observed using the {\it Ooty Radio Telescope} (ORT), 
{\it Solar-Terrestrial Environment Laboratory} (STEL) and {\it Mexican Array Radio Telescope} 
(MEXART), providing insight on solar-wind properties \citep{man90,man94,jul10}.

Solar events can affect the plasma of the Earth's ionosphere either directly, through X-ray 
and/or EUV radiation from flares as it reaches Earth, or indirectly, through the perturbation of 
the magnetospheric electric field upon the arrival of an interplanetary CME \citep{Tsu09}. The 
Earth's ionosphere is considered a dispersive medium for radio waves. Its refractive index is a 
function of the radio-wave frequency, the electron density, and the intensity of the Earth's magnetic 
field, {\bf B}. Its perturbation can be an important source of error for the signals of the {\it Navstar 
Global Positioning System} (GPS) satellites and other positioning systems like {\it Galileo} 
(European Union), {\it Glonass} (Russia), and {\it Compass} (China). The error in the signals 
is proportional to the integrated electron density along the signal path, TEC, and inversely 
proportional to the square of the carrier phase frequency, $\tau \propto TEC/\nu ^2$ 
\citep{lan96,kom97}. 
The signals from GPS satellites must travel through the Earth's ionosphere in their way to the GPS 
receivers (on the Earth's surface) in the L-band: $L_1=1575.42$\,MHz and $L_2=1227.60$\,MHz. 
 Although there is another L-band, $L_3=1381.05$\,MHz, GPS users cannot use it \citep{eri01}.

Several works have shown the importance of the ionospheric interference on the IPS \citep{gapp82,pur87,tap84,woa95,luc96,jac98,per08,shi10}. Part of this contamination, or 
ionospheric scintillation (IONS), can represent a serious problem in IPS studies. In fact, the 
presence of IONS can lead to a misinterpretation of the IPS; however, there are techniques to 
remove the ionospheric scintillation part. \citet{car11} have explored another possible 
data contamination by ionospheric effects. They have shown an example related to a solar event 
on 15 December 2006, where the fluctuations observed in the signal of the radio sources exhibited 
strong contamination by ionospheric disturbances (IOND).  The behavior shown by the radio signal 
was consistent with the fluctuations observed from the signal emitted by polar satellites. The 
authors suggested that not only the total electron content (TEC) and the geomagnetic equatorial 
index, {\it Dst}, could be used to characterize the origin of such fluctuations, but also the possibility 
of removing the contaminating frequencies from the data to study separately the IPS or the IOND. 

Since we have observed fluctuations in the radio signal of sources transiting the MEXART 
observatory, the purpose of our study is to understand the nature of these fluctuations, as well as 
answering the following questions: Are the fluctuations intrinsic to quiet solar-wind conditions? 
When are these fluctuations related to ionospheric disturbances? How do these fluctuations 
quantitatively affect the IPS? To answer these questions, we have analyzed the radio signal of 
sources observed with the MEXART. We have explored all observations to find if the 
ionospheric contribution is always present, even in the absence of solar events. In Section 
\ref{obs}, we present the information related to the MEXART observations for the period between 
April and May 2010. Section \ref{dat_an} shows the data analysis applied and the results using  
wavelet analysis applied to the MEXART data. Finally, in Sections \ref{dis} and \ref{con} we 
present the discussion and conclusions of our work, respectively.

\section{Observations}
\label{obs}
%Text with citations \cite{RefB} and \cite{RefJ}.

We considered two short periods of MEXART (located at a latitude of $+19^{\circ}$ $48'$  $39''$, 
longitude of $-101^{\circ}$ $41'$ $39''$, and elevation of 1964\,m above mean sea level) 
observations, from 20 to 30 April and 8 to 
31 May 2010, with a configuration that used a small section of the MEXART antenna. We used 
a 1024-dipole rectangular sub-array arranged in 16 rows. Each row is horizontally polarized in the 
East-West direction. The total area of the antenna is 2415 square meters, 17.25\,m (North-South) 
$\times$ 140\,m (East-West), which is one fourth of the total MEXART array. The array feeds a 
16x16 Butler matrix, which generates 16 beams, each of width 1x8 degrees. The receiver works at 
139.65 MHz with a 2 MHz bandwidth and a time constant of 40 ms. The acquisition rate 
is 20 ms. Further technical details of the radio telescope can be found elsewhere 
\citep{gonz04,gonz06,car07}. The daily MEXART observations and the transit of strong radio 
sources can be viewed in real time\footnote{http://www.mexart.unam.mx}.

\begin{figure}
\centering%
\includegraphics[width=7.0cm,angle=90]{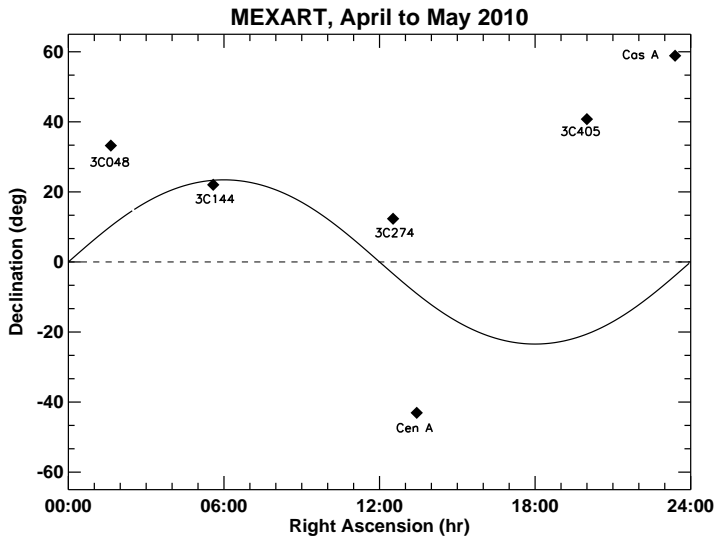}
\caption{Location of the radio sources observed with MEXART. These radio sources 
were observed between 20 April and 31 May 2010. The orbit 
of the Sun along the year is indicated with a  
continuous line.}
\label{fig1}
\end{figure}

Transit observations of the following sources were taken on several days: 
3C48 (R.\,A.\,$=01^h 37^m 41.3^s$, Dec\,$= +33^{\circ} 09^m 35^s$, right ascension and  declination respectively), 
3C144 (R.\,A.\,$=05^h$ $34^m$ $32.0^s$, Dec\,$= +22^{\circ}$ $00^m$ $52^s$), 
3C274 (R.\,A.\,$=12^h$ $30^m$ $49.4^s$, Dec\,$= +12^{\circ}$ $23^m$ $28^s$), 
Cas A (R.\,A.\,$=23^h$ $23^m$ $27.9^s$, Dec\,$= +58^{\circ}$ $48^m$ $42^s$), 
Cen A (R.\,A.\,$=13^h$ $25^m$ $27.6^s$, Dec\,$= -43^{\circ}$ $01^m$ $09^s$), 
and 
3C405 (R.\,A.\,$=19^h$ $59^m$ $28.3^s$, Dec\,$= +40^{\circ}$ $44^m$ $02^s$). 
The coordinates are precessing at equinox J2000.0 and are represented in a map 
(Figure \ref{fig1}) showing the right ascension (R.\,A.) and declination (Dec).  Table 
\ref{table1} lists the observation dates of the sources whose radio signal presented fluctuations   
when transiting the MEXART observatory (see first and second columns). We use the MEXART 
position (latitude, longitude and elevation) and the  precessed coordinates (R.\,A. and Dec) for each 
radio source to determine their transit time during the date we analyze MEXART data. The third 
column in Table \ref{table1} shows the corresponding elongation angle ($\epsilon$, the angle 
between the Sun-Earth-line and the source) in degrees for each source. The radio source 3C144 
shows the smallest $\epsilon$ values, all fluctuations shown by these sources were analyzed to 
characterize their origin. There are days with no MEXART observations; they appear as blank in 
Table \ref{table1}.

\subsection{{The \it vTEC} and the {\it Dst} index}
\label{tec_dstindex}

When analyzing radio sources observed by any telescope, such as MEXART, it is important to take 
into account the conditions in the environment. The origin of the fluctuations observed in the radio 
signal of each source can be understood if we analyze them together with the characteristics of the 
geomagnetic field and the ionosphere. In this context, we included the geomagnetic equatorial {\it 
Dst} index (the provisional values from 
Kyoto\footnote{\url{http://wdc.kugi.kyoto-u.ac.jp/dstdir/}}) and the vertical total 
electron content\footnote{Usually the {\it vTEC} is measured in {\it TEC} units ({\it TECu}), where 
$1$\,$TECu=10^{16}$\,m$^{-2}$.} ({\it vTEC}) to understand the potential correlation 
between geomagnetic storms and ionospheric disturbances. The 
ionospheric total electron content is a well suited parameter to study the conditions of the perturbed 
ionosphere and it is particularly important to correct the positioning information for single-frequency 
GPS users \citep{ara05,ara06}. To facilitate the comparison between the {\it Dst} index and the {\it 
vTEC} properties, in the fourth and fifth column of Table \ref{table1}, we list the averaged {\it Dst} 
index, $<Dst>$, and the corresponding minimum value of {\it Dst}, $Dst_{min}$, for each day. 
Finally, we have included the dispersion of  {\it vTEC}, $\delta $, which is defined as:

\begin{equation}
\delta = \frac{(vTEC_{max}-<vTEC>)}{<vTEC>},
\end{equation}
  
\noindent values of $\delta \gtrsim 4$ indicate that {\it vTEC} is strongly 
disturbed \citep[see for instance][]{lop12,car11}.

%%%%%%%

\begin{longtable}{llllll}
\caption{Sources showing fluctuations in the radio signal observed by MEXART observatory. The 
$\delta$ parameter is the dispersion of {\it vTEC} (see Section 2.1).}\\
%{\footnotesize\tt
%\begin{tabular}{llllll}
\hline \hline
Date   & Sources & $\epsilon$   & $<Dst>$  & $Dst_{min}$ & $\delta$ \\
(2010) &         & ($^{\circ}$) & (nT)     & (nT)        &          \\   
\hline
20 April &                     &  &  1  & -10 & 1.50 \\
21 April &                     &  &  1  & -7  & 1.33 \\
22 April &                     &  &  5  & -5  & 1.71 \\
23 April & 3C144, 3C405        & 50.6, 82.6 &  -9 & -22 & 1.48 \\
24 April & Cas A               & 55.1 & -12 & -23 & 1.47 \\
25 April & 3C144, 3C405, Cas A & 48.7, 83.6, 55.2 &  1  & -5  & 1.65 \\
26 April & 3C274, Cen A, Cas A & 143.7, 148.8, 55.3 &  7  &  2  & 1.23 \\
27 April & 3C144, Cas A        & 46.7, 55.4 &  5  &  0  & 1.33 \\
28 April & 3C144, 3C405, Cas A & 45.8, 85.1, 55.6 &  7  & -3  & 1.56 \\
29 April & 3C405, Cas A        & 85.6, 55.7 &  2  & -6  & 1.55 \\
30 April &                     &  &  0  & -4  & 1.75 \\
%May 1    & No MEXART Data      &  &  3  & -8  & 1.05 \\
%May 2$^{nd}$    & No MEXART Data      &  & -50 & -67 & 0.98 \\
%May 3    & No MEXART Data      &  & -28 & -53 & 1.17 \\
%May 4    & No MEXART Data      &  & -23 & -33 & 1.25 \\
%May 5    & No MEXART Data      &  & -13 & -23 & 1.38 \\
%May 6    & No MEXART Data      &  & -15 & -23 & 1.77 \\
7 May    &                     &  & -14 & -23 & 1.35 \\
8 May     & 3C144               & 36.1 & -8  & -28 & 1.10 \\
9 May    &                     &  & -6  & -13 & 1.41 \\
10 May    &                     &  &  4  & -8  & 1.38 \\
11 May   & 3C274, Cen A        & 130.2, 146.4 &  8  & -8  & 1.76 \\
12 May   &                     &  &  -8 & -12 & 1.09 \\
13 May   &                     &  &  1  & -10 & 1.46 \\
14 May   &                     &  &  3  & -3  & 1.97 \\
15 May   &                     &  &  -3 & -13 & 1.92 \\
16 May   &                     &  &  4  & -3  & 1.72 \\
17 May   &                     &  &  -1 & -12 & 2.29 \\
18 May   &                     &  &  -3 & -28 & 2.37 \\
19 May   & Cas A               & 60.4 &  0  & -2  & 1.93 \\
20 May   & 3C405, Cas A        & 95.8, 60.7 & -12 & -17 & 1.64 \\
21 May   & 3C405               & 96.3 &  3  & -11 & 1.96 \\
22 May   & 3C405, Cas A        & 96.8, 61.4 &  3  & -3  & 1.43 \\
23 May   & 3C405, Cen A        & 97.3, 141.0 &  5  & -1  & 1.16 \\
24 May   & 3C405               & 97.7 &  10 &  0  & 1.17 \\
25 May   & 3C405               & 98.2 &  9  &  0  & 1.34 \\
26 May   & 3C405, Cas A        & 98.7, 62.9  & -4  & -4  & 1.57 \\
27 May   & Cas A               & 63.3 & 1   & -3  & 1.45 \\
28 May   & 3C144, 3C405, Cas A & 16.7, 99.6, 63.7  & -3  & -3  & 1.52 \\
29 May   & 3C144, 3C405, Cas A & 16.0, 100.1, 64.1 & -29 & -85 & 1.28 \\
30 May   &                     &  & -37 & -48 & 1.93 \\
May 31   &                     &  & -32 & -37 & 1.34 \\
\hline
%\end{tabular} 
%}
\label{table1}
\end{longtable}

%%%%%%%%%%%%%%%%%%%%%%

The {\it vTEC} values reported here were calculated using five GPS 
receivers in stations that cover a wide region of the sky and that were located in the proximity of the  MEXART observatory: Coeneo (UCOE), Celaya (CEGA), Aguascalientes (INEG), Oaxaca (OAX2) 
and Toluca (TOL2). Table \ref{table2} shows the location (latitude and longitude) 
and the elevation of the GPS-signal recording stations.

\begin{table}
\caption{Location of the GPS-signal recording stations used to calculate the {\it vTEC}. The fifth column corresponds to the elevation above sea level of each station.}
\begin{tabular}{lllll}
\hline \hline
Monument Name & Monument Code & Latitude (N) & Longitude (W) & Elevation (m) \\  
\hline 
Coeneo & UCOE & 19$^{\circ}$ 48$^{'}$ 47$^{''}$ & 101$^{\circ}$ 41$^{'}$ 39$^{''}$ & 1978.8 \\
Celaya & CEGA & 20$^{\circ}$ 31$^{'}$ 40$^{''}$ & 100$^{\circ}$ 48$^{'}$ 55$^{''}$ & 1750.0 \\
Aguascalientes & INEG & 21$^{\circ}$ 51$^{'}$ 22$^{''}$ & 102$^{\circ}$ 17$^{'}$ 03$^{''}$ & 1888.4 \\
Oaxaca & OAX2 & 17$^{\circ}$ 04$^{'}$ 42$^{''}$ & 96$^{\circ}$ 43$^{'}$ 00$^{''}$ & 1607.3 \\
Toluca & TOL2 & 19$^{\circ}$ 37$^{'}$ 35$^{''}$ & 99$^{\circ}$ 38$^{'}$ 36$^{''}$ & 2651.9 \\
\hline
\end{tabular}
\label{table2}
\end{table}

\section{Data Analysis and Results}
\label{dat_an}

The data analysis was done in two steps: first we analyze MEXART data and second we  study the ionospheric data. 

\subsection{MEXART Data}
\label{mex_dat}

Data analysis of MEXART radio signals yielded two statistical parameters: the signal to 
noise ratio (S/N) and the average root mean square (rms) value calculated for the previous and the ensuing days for each radio source observational period. We analyzed the time series for 
each source transit-time (around 8 minutes), looking for strong fluctuations in the radio signals.  

We used several routines developed to analyze MEXART data in which we included one to read the raw data, another one to remove interference and gaps, and a third one that included gaussian fits to obtain a detrended time series.

\begin{figure}
\centering%
\includegraphics[width=6.0cm,angle=270]{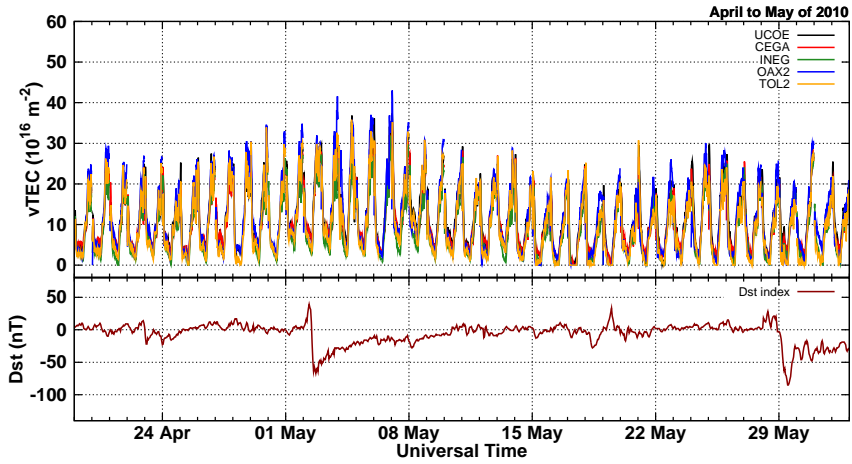}
\caption{The vertical total electron content ({\it vTEC}) and {\it Dst} index. The top panel 
shows the {\it vTEC} calculated from five stations: UCOE, CEGA, INEG, OAX2 and TOL2, represented with lines in black, red, 
green, blue and orange colors respectively (see section \ref{tec_dstindex}). The bottom panel shows the {\it Dst} 
index. Both, {\it vTEC} and {\it Dst} index, are shown in the time interval from April to 
May 2010. }
\label{fig2}
\end{figure}

The time interval covered by the GPS data was 
the same, 20 April to 31 May, 2010. During this period, 
the GPS data showed the typical diurnal variation without any 
significant event. However, the {\it vTEC} showed a maximum of about 30\,{\it TECu} 
associated with the geomagnetic storm with a minimum Dst of -85 nT,  which occurred on 29 May. 
Figure \ref{fig2} shows the {\it vTEC} for each station, where the {\it vTEC} series is given every 15 
minutes. The lower panel in Figure  \ref{fig2} shows the evolution of the {\it Dst} index. We observe 
two minima, one around 2 May (with a minimum value of -67 nT at 18:00 UT) and a second one on 
29 May (with a minimum value of -85 nT between 13:00 and 14:00 UT).

The two minima, mentioned above, were probably associated with the coronal mass ejections (CME\footnote{See the Large Angle and Spectroscopic (LASCO) CME catalog at 
 \url{http://cdaw.gsfc.nasa.gov/CME_list/}}) 
that occurred in active regions 11063 and 11072, or with the coronal holes located close to central meridian around four days before they were registered. These solar events certainly 
modified the plasma conditions in the interplanetary medium, and possibly along the
line of sight between the radio sources and the MEXART observatory.

Four slow and narrow CMEs were observed on 29 April, which could be associated with the minimum on 2 May; however, these CMEs seemed to have no significant effect on the Earth's  magnetosphere.  The CMEs occurred at 01:12 UT, 03:36 UT, 07:00 UT, and 17:00 UT, and had velocities of 138 km s$^{-1}$, 178 km s$^{-1}$, 187 km s$^{-1}$, and 378 km s$^{-1}$, respectively. Their position angles (PAs) were around 80$^{\circ}$ for the first three, and 260$^{\circ}$ for the last one. All these CMEs seemed to originate from active region 11063 (N16 E11), very close to central meridian.

A coronal hole was observed in X-ray images obtained with satellite {\it GOES} 14 by the on of 24 April. This coronal hole appeared as a small region at the pole to the North--East of the solar disk. The coronal hole displaced towards the North--West during several days  and reached lower latitudes where it became bigger. The high speed streams ($\sim$600 to 700 km\,s$^{-1}$) originating in the coronal hole were registered by the plasma instruments onboard {\it Wind} and the 
{\it Advanced Composition Explorer} (ACE) from 30 April to 2 May 2010.  Before the initiation of the geomagnetic storm, the solar-wind speed had an average value of $\sim$400 km\,s$^{-1}$. However, on 2 May, this value increased to around 650\,km\,s$^{-1}$. This fact coincides with the growth of the plasma temperature (reaching $7\times 10^5$\,K). The magnetic field magnitude also showed an increase, which is consistent with an observed rotation of the southward component of the magnetic field, B$_z$ and with the growth of the plasma pressure. Finally, the plasma $beta$ parameter exhibited  small values (close to zero) in the same time interval, confirming the effect of the high speed stream from this coronal hole.

Regarding the second minimum in the Dst index on 29 May, the active region 11072 (S15 W11) displayed  some activity during several days starting on 24 May. A halo CME, with a linear speed of 427\,km\,s$^{-1}$ occurred on 24 May at $\approx$ 14:06 UT.  X-ray images from {\it GOES} 14 also showed a coronal hole by the end of 24 May; it appeared as a small region at the pole to the North--East of the solar disk. This structure continuously moved to lower latitudes and became bigger. Thus, the presence of the active region, the CMEs occurrence, and the high speed stream from a coronal hole, which displayed the characteristics of a magnetic cloud (with a nice $B_z$ rotation), may have led to the dip in {\it Dst} index. The {\it vTEC} showed a slight response to these events, whereas the {\it Dst} index presented minima. The total electron content had a maximum no greater than 45\,{\it TECu} between 3 to 6 May, and no greater than 35\,{\it TECu} between 29 and 31 May. Unfortunately, we do not 
have MEXART data for the period of 1 to 7 May. Depending on the 
availability of MEXART observations,  we estimated the scintillation index for each source as in \citet{per06}.

\subsection{Wavelet Analysis}
\label{wavelets}

A wavelet transform can map the power of a particular frequency at different 
times, giving an expansion of the signal in both time and frequency. Furthermore, the wavelet transform not only tells us which frequencies can exist in 
the signal, but also shows whether a 
particular scale varies in time. In addition, the wavelet tool has an important advantage 
over Fourier transforms since it can extract frequency information from a 
signal using search windows of variable scales. 

The wavelet analysis used in this work is based mainly on the wavelet 
software provided by C. Torrence and G. 
Compo\footnote{http://atoc.colorado.edu/research/wavelets/}, and  programmed in the Interactive Data Language (IDL). We also included several new 
routines adapted to this software for analyzing MEXART data. These routines were 
developed to read the data and fitting gaussian functions to get 
a detrended time-series, which was used in our wavelet approach.

The wavelet analysis was applied to the radio signal of every source day by day 
to look for fluctuations that stand above the background signal along 
the time interval for the data analysis used in this work. With this 
tool we characterized the frequencies associated with these fluctuations 
in perturbed 
time subintervals (where the {\it Dst} geomagnetic index and TEC were slightly 
perturbed) and  we compared them with undisturbed subintervals.

\begin{figure}
 \includegraphics[width=11.5cm,angle=0]{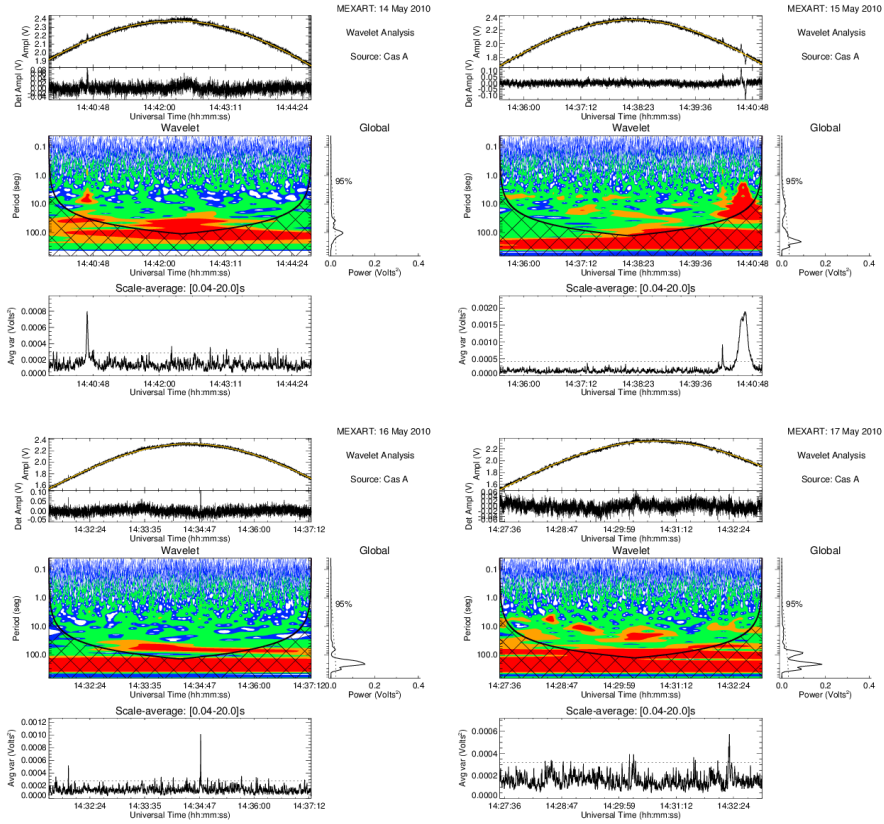}
 \caption{Wavelet analysis for the 
 radio source Cas A, from 14 to 17 May. The radio signal of this source does not show significant 
 disturbances during these days. These are characteristic wavelet-spectra of 
 quiet days for this particular source. The level colors (black to red) for the 
 power of each spectrum were:  [0.0001, 0.0006, 0.007, 0.02, 0.5, 0.8, 2.0]\,V$^2$. See Section 3.2 
 for the explanation of every panel in this figure.}
 \label{fig3}
 \end{figure}

\begin{figure}
 \includegraphics[width=11.5cm,angle=0]{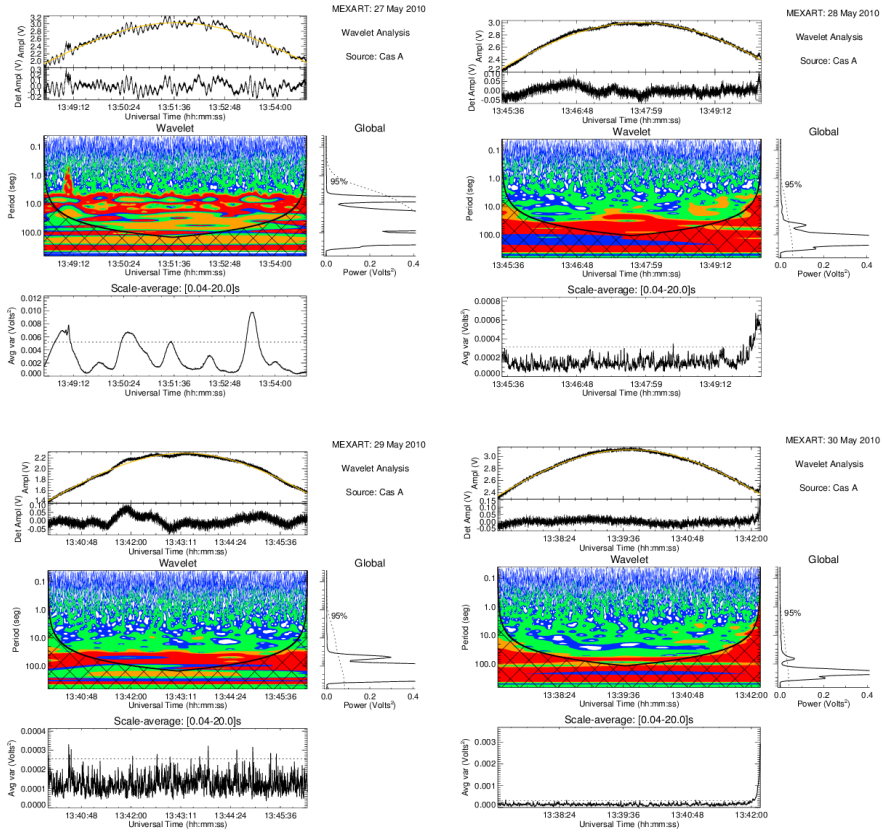}
 \caption{Wavelet analysis for the 
 radio source Cas A, from 27 to 30 May. The fluctuations observed on 27 May could be 
 associated with perturbations in the IPM producing IPS and, similarly, or 28 to 29 May. 
 This source shows slight fluctuations in the signal related to a CME-Halo event (see Section 3). 
 The level colors are the same as in Figure \ref{fig3}. See Section 3.2 for the explanation of every 
 panel in this figure.}
 \label{fig4}
 \end{figure}

\begin{figure}
 \includegraphics[width=11.5cm,angle=0]{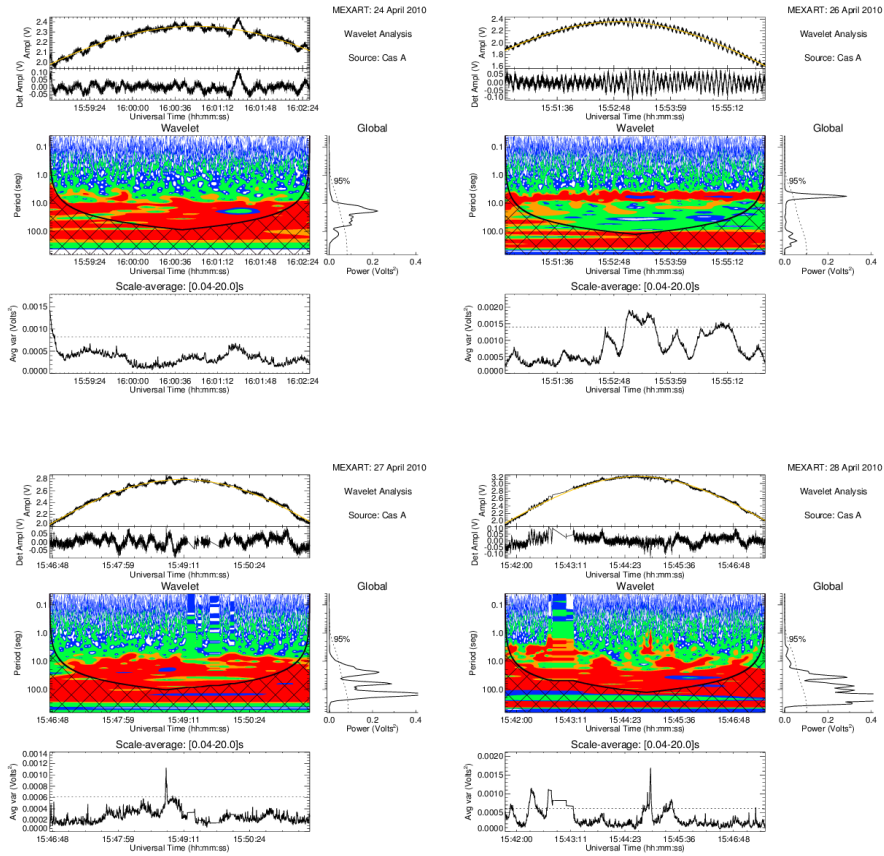}
 \caption{Wavelet analysis for the 
 radio source Cas A from 24 to 28 April. For all of these days, Cas A shows clear and different 
 fluctuations. On 24 and 26 April, these fluctuations may be due to perturbations in the IPM 
 producing IPS; however, on 27 and 28 April, the fluctuations may be associated with IPS and 
 IOND. The level colors are the same as in Figure \ref{fig3}.  See Section 3.2 for the explanation of 
 every panel in this figure.}
 \label{fig5}
 \end{figure}

\begin{figure}
 \includegraphics[width=11.5cm,angle=0]{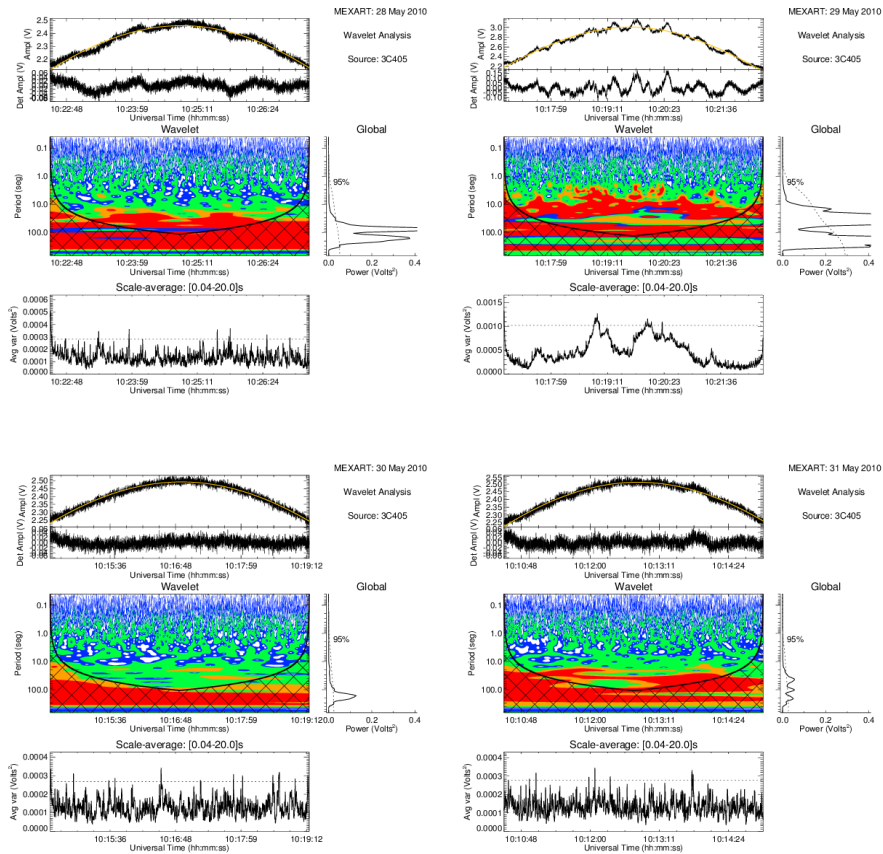}
 \caption{Wavelet analysis for the 
 radio source 3C405 from 28 to 31 May. This source shows strong fluctuations on 29 May, probably 
 associated with IPS (see Section 4). The level colors are the same as in Figure  \ref{fig3}. See 
 Section 3.2 for the explanation of every panel in this figure.}
 \label{fig6}
 \end{figure}

\begin{figure}
 \includegraphics[width=11.5cm,angle=0]{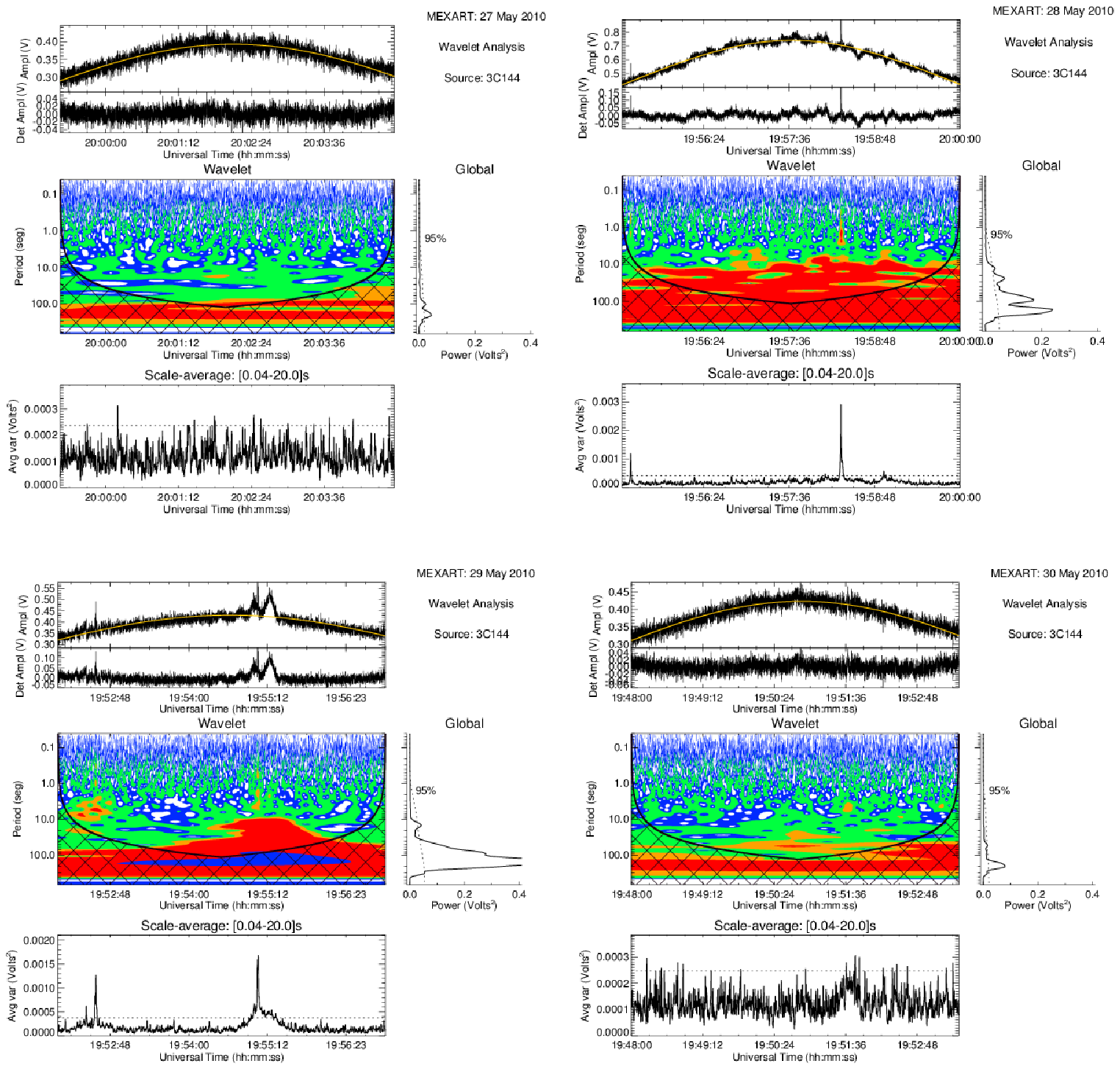}
 \caption{Wavelet analysis for the 
 radio source 3C144, from 27 to 30 May (see Section 4). The level colors are the same as in Figure 
 \ref{fig3}. See Section 3.2 for the explanation of every panel in this figure.}
 \label{fig7}
 \end{figure}

\begin{figure}
 \includegraphics[width=11.5cm,angle=0]{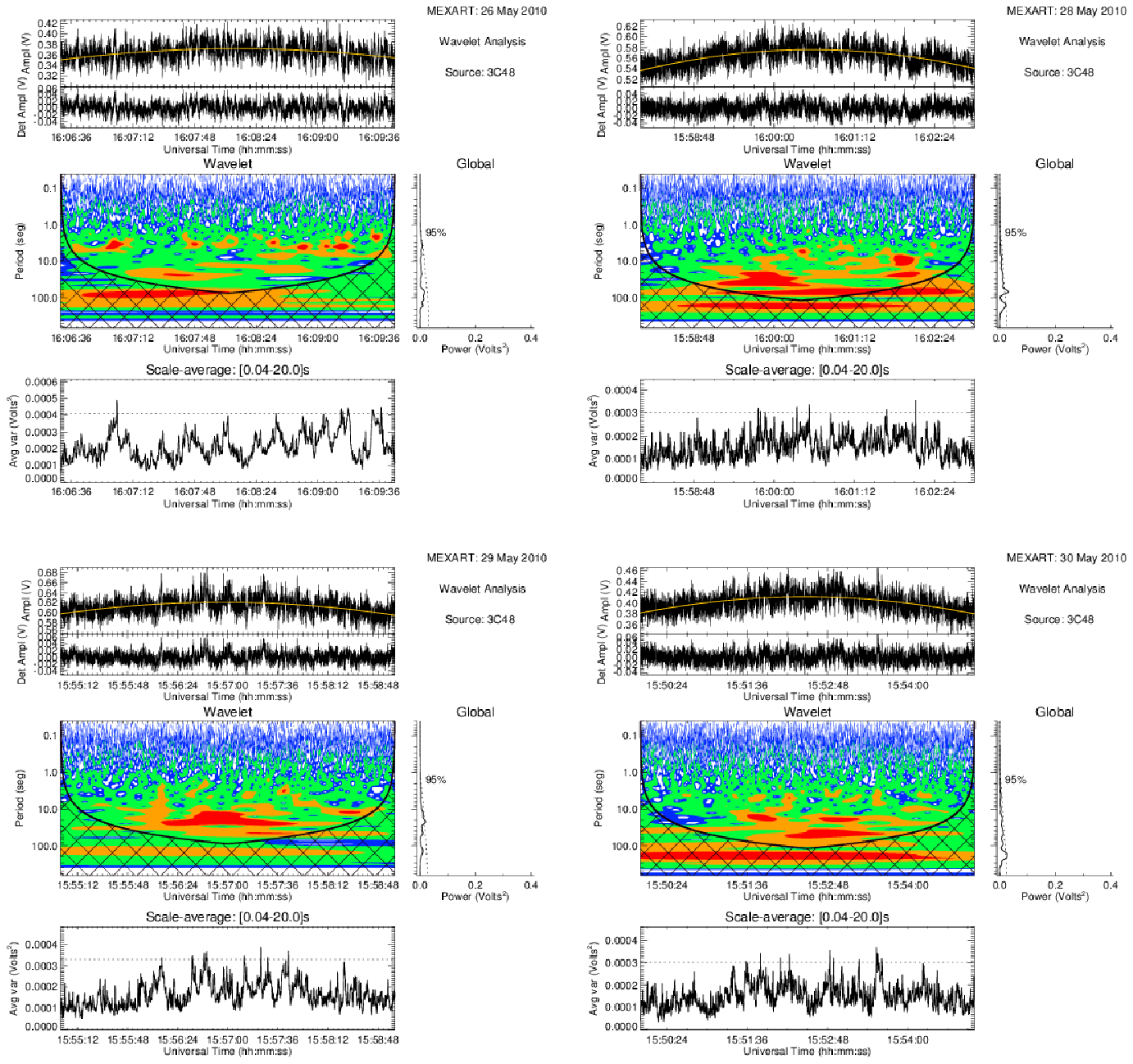}
 \caption{Wavelet analysis for the 
 radio source 3C48 from 26 to 30 May (see Section 4). The level colors are the same as in Figure 
 \ref{fig3}. See Section 3.2 for the explanation of every panel in this figure.}
 \label{fig8}
 \end{figure}

Figures \ref{fig3} to \ref{fig8} 
show wavelet plots for representative dates and sources. Each figure considers 4-day observations for one source. From top to bottom and from left to right we present in the 
first two panels, the time series together with a gaussian fitting (thick yellow line)
and the amplitude of the detrended time-series below. In addition, the middle panels show the 
wavelet of the detrended time-series depicting the periodicities associated with the observations. The black lines indicate 
the influence of the cone region at a 95\% significance level. To the right,  
the wavelet window shows the spectral power-spectrum highlighting the 
periodicities/frequencies above a line which represents the 95\% significance 
level. Finally, the bottom panel shows the average variance.

Table \ref{table3} shows the best gaussian-fit parameters applied to the radio data  
in order to obtain a detrended signal. We selected representative 
dates for perturbed ({\it i.e.} 24 to 31 May) and unperturbed ({\it i.e.} 24 to 30 April  and  13 to 18 May time 
intervals) TEC for every source that reflects this behavior in several parameters. 
Table \ref{table3} also contains signal to noise ratio (S/N) values in the 
second column, the amplitude (A) in volts in the third column, the transit time 
(T.T.) recorded by the MEXART observatory for every source, the full-width at half-maximum
(FWHM) in minutes, and the standard error between the fit and the radio signal, 
shown in the fourth, fifth, and sixth columns, respectively. The S/N ratio 
for the sources presented in this work took values between 4.1 to 36.8 (see 
Table \ref{table3}), showing that the parameters obtained in our analysis were well determined. 

%%%%%%%%%%%%%%%%%%%%%%%%%%%%%%%%

\begin{longtable}{lccccc}
\caption{Parameters for the best fit to the radio signal obtained from the wavelet analysis, where 
S/N is the signal to noise ratio, A is the amplitude for each radio signal, T.T. is the transit time for each source, FWHM is the full-width at half-maximum, and $E_{std}$ is the standard error between the fit and the signal-data.}\\
%\begin{tabular}
\hline \hline
Date & {S/N} & {A (Volts)} & {T.T. (hr)} & {FWHM (min)} & {$E_{std}$} \\
\hline  
\multicolumn{6}{c}{Source: Cas A} \\
24 Apr. 2010 & 22.0 & 2.35 & 16.01 &  8.23 & 0.03 \\
26 Apr. 2010 & 10.4 & 2.36 & 15.88 &  8.33 & 0.03 \\
27 Apr. 2010 & 11.1 & 2.79 & 15.82 &  7.15 & 0.03 \\
28 Apr. 2010 &  7.0 & 3.18 & 15.74 &  6.92 & 0.03 \\
14 May 2010 & 14.3 & 2.38 & 14.70 &  8.13 & 0.01 \\
15 May 2010 & 10.6 & 2.34 & 14.64 &  8.03 & 0.02 \\
16 May 2010 &  9.4 & 2.32 & 14.58 &  8.04 & 0.01 \\
17 May 2010 &  9.4 & 2.33 & 14.51 &  8.23 & 0.02 \\
\multicolumn{6}{c}{Source: 3C405} \\
25 Apr. 2010 & 19.5 & 3.07 & 12.55 &  8.03 & 0.01 \\
26 Apr. 2010 & 14.8 & 2.89 & 12.49 &  8.23 & 0.02 \\
27 Apr. 2010 & 29.6 & 1.12 & 12.43 & 10.25 & 0.01 \\
28 Apr. 2010 &  9.0 & 3.47 & 12.37 &  6.04 & 0.02 \\
28 May 2010 & 22.8 & 2.46 & 10.41 & 10.34 & 0.02 \\
29 May 2010 & 10.1 & 3.00 & 10.33 &  7.89 & 0.05 \\
30 May 2010 & 30.6 & 2.49 & 10.28 & 11.98 & 0.01 \\
31 May 2010 & 29.4 & 2.51 & 10.21 & 11.17 & 0.01 \\
\multicolumn{6}{c}{Source: 3C144} \\
22 Apr. 2010 & 12.5 & 0.93 & 22.32 &  5.68 & 0.01 \\
23 Apr. 2010 & 12.4 & 0.93 & 22.26 &  5.61 & 0.01 \\
24 Apr. 2010 &  9.2 & 0.79 & 22.19 &  5.53 & 0.01 \\
25 Apr. 2010 &  8.4 & 0.86 & 22.12 &  5.85 & 0.01 \\
28 May 2010 &  6.2 & 0.74 & 19.96 &  5.26 & 0.02 \\
29 May 2010 & 10.4 & 0.43 & 19.91 &  8.04 & 0.02 \\
30 May 2010 & 11.1 & 0.42 & 19.85 &  8.54 & 0.01 \\
31 May 2010 &  8.9 & 0.79 & 19.77 &  5.41 & 0.02 \\
\multicolumn{6}{c}{Source: 3C48} \\
22 Apr. 2010 & 33.9 & 0.69 & 18.39 & 21.81 & 0.02 \\
24 Apr. 2010 & 27.1 & 0.58 & 18.24 & 16.56 & 0.02 \\
25 Apr. 2010 & 27.0 & 0.55 & 18.17 & 16.80 & 0.02 \\
26 Apr. 2010 & 31.9 & 0.61 & 18.12 & 17.22 & 0.02 \\
26 May 2010 & 21.2 & 0.37 & 16.14 & 11.69 & 0.02 \\
28 May 2010 & 30.7 & 0.58 & 16.01 & 16.09 & 0.01 \\
29 May 2010 & 36.8 & 0.62 & 15.95 & 16.63 & 0.01 \\
30 May 2010 & 23.4 & 0.41 & 15.87 & 14.86 & 0.01 \\
\multicolumn{6}{c}{Source: 3C274} \\
23 Apr. 2010 &  4.1 & 0.27 &  5.23 &  4.51 & 0.01 \\
24 Apr. 2010 &  5.5 & 0.30 &  5.16 &  5.48 & 0.01 \\
25 Apr. 2010 &  4.7 & 0.24 &  5.10 &  4.65 & 0.01 \\
26 Apr. 2010 &  6.8 & 0.26 &  5.03 &  4.19 & 0.01 \\
14 May 2010 &  4.8 & 0.27 &  3.85 &  4.66 & 0.01 \\
15 May 2010 &  7.5 & 0.26 &  3.79 &  4.13 & 0.01 \\
16 May 2010 &  5.1 & 0.25 &  3.72 &  4.42 & 0.01 \\
17 May 2010 &  9.2 & 0.27 &  3.65 &  4.51 & 0.02 \\
\multicolumn{6}{c}{Source: Cen A} \\
23 Abr. 2010 &  9.2 & 0.91 &  6.14 &  7.44 & 0.03 \\
24 Abr. 2010 & 13.8 & 0.92 &  6.07 &  7.66 & 0.02 \\
25 Abr. 2010 & 14.2 & 0.88 &  6.00 &  7.41 & 0.01 \\
26 Abr. 2010 & 15.0 & 0.86 &  5.94 &  7.11 & 0.01 \\
28 May 2010 & 14.3 & 0.78 &  3.84 &  7.16 & 0.02 \\
29 May 2010 &  9.7 & 0.70 &  3.78 &  6.35 & 0.01 \\
30 May 2010 & 10.2 & 0.85 &  3.71 &  7.44 & 0.03 \\
31 May 2010 & 15.1 & 0.78 &  3.64 &  7.19 & 0.03 \\
\hline
%\end{tabular}
\label{table3}
\end{longtable}

%%%%%%%%%%%%%%%%%%%%%%%%%%%%%%%

\section{Discussion}
\label{dis}

\begin{figure}
 \begin{center}
 \includegraphics[width=7.5cm,angle=0]{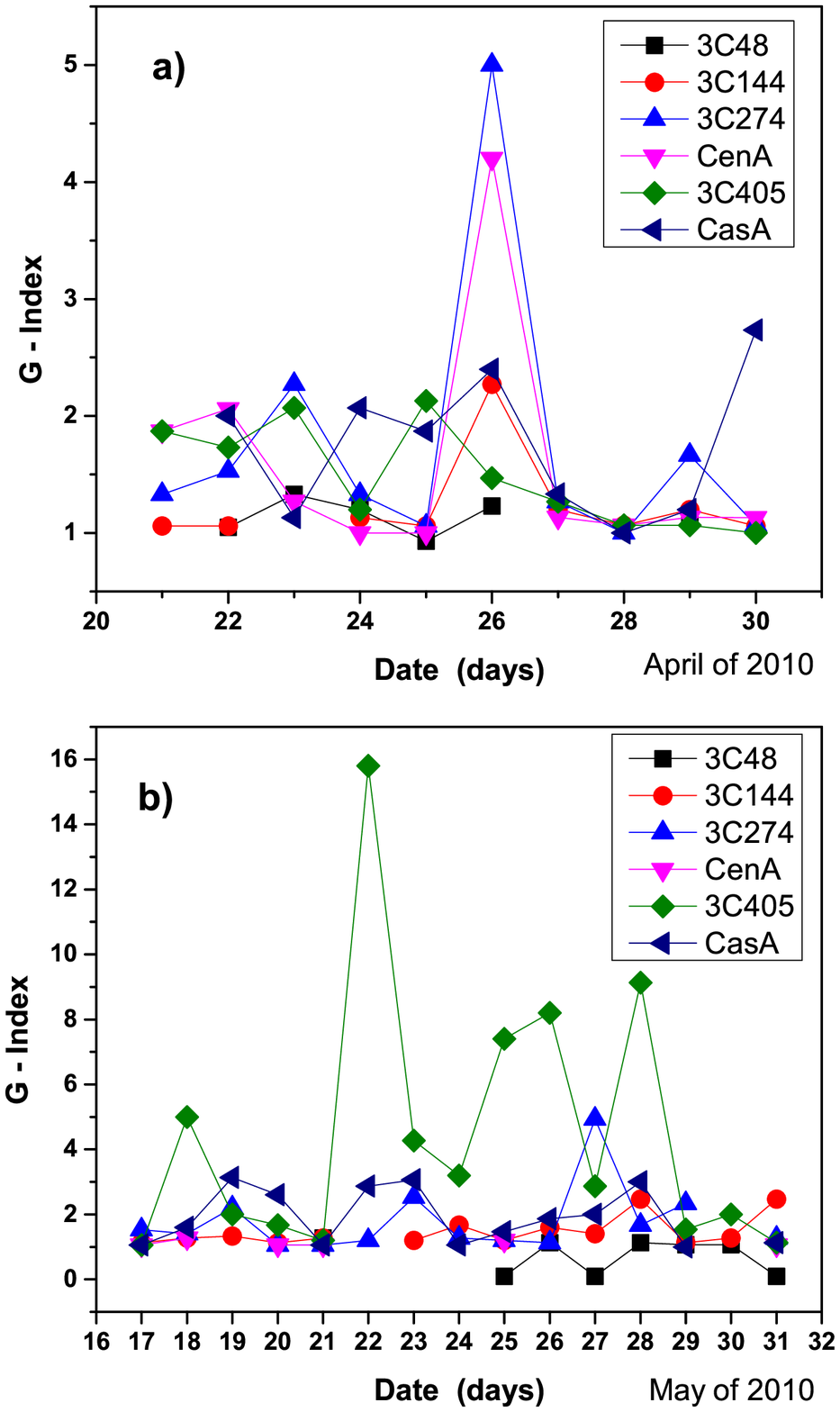}
 \caption{The G-index, as defined in \citet{per06}, is plotted as a function of time 
 (in days). Panel a) shows the behavior for each radio source used in our analysis from 21 to 30  
 of April, close to the time of the first {\it Dst} minimum  on 2 May. On 26 April, 3C274 
 and Cen A exhibit a significative G-index, its value was twice large than for others sources. Panel 
 b) shows 
 the same data as panel a), but from 16 to 31 May. The source 3C405 shows 
 several peaks indicating a high G-index along the line of sight (see 
 the text for more details in section 4).}
 \label{fig9}
 \end{center}
 \end{figure}

In this paper we compared the perturbations observed with the MEXART radio telescope and those in the {\it vTEC} and the {\it Dst} index for the periods of observation from 20 to 30 April and from 8 to 31 May. Several sources were found to scintillate and we analyzed in which cases the signals were contaminated by IOND (as was found by \citet{car11}). 
Figure \ref{fig9} shows the G index computed as: 

\begin{equation}\label{eq-g}
G = \left( \frac{\langle \Delta I(t)^2\rangle}{\langle I(t)\rangle^2} \right)^{1/2}, 
\end{equation}

\noindent for each source plotted as a function of time (in days), where the source intensity at a given time is $I(t)$ and its fluctuation around the mean is $\Delta I(t)$. The fluctuation is defined as 
$\Delta I(t)=I(t)-\langle I(t)\rangle$ and the mean intensity of the source is defined as 
$I_0\sim\langle I(t)\rangle$ \citep{per06}. From  panel a), we observe that the G index shows an increase on 26 April for the sources Cen A and 3C274. Since the {\it Dst} index and the {\it vTEC} correspond to quiet periods, this suggests that these 
fluctuations are probably associated with perturbations in the interplanetary medium producing IPS. Moreover, Cas A also presents an increase on 30 April, suggesting an ionospheric origin.  From panel b),  we observe along this time interval that the G index for source 3C405 has stronger peaks.  Due to the position of this source, we consider that its radio-signal behavior is related to ionospheric disturbances around the auroral region.

We discuss the behavior of the signal of each source: 

\begin{enumerate}
\item {\it Cas A} shows no significant fluctuations in an undisturbed period 
	from 14 to 17 May, as shown in Figure 
	\ref{fig3}. The behavior of the radio signal is very similar for 
	several days. Additionally, the wavelet analysis shows no significant 
	periodicities/frequencies within the cone of influence  for 
	each day. This is in agreement with the observed power spectra, 
	which show similar behaviors with no significant periodicities 
	beyond a 95\% significance level. Furthermore, we observe that 
	sometimes there are narrow peaks mounted on the source radio-signal that modify the 
	variance substantially
	(bottom panel). These peaks are associated with lightning storms close 
	to the MEXART observatory and/or with transiting-satellite signals. We 
	assume that these signals do not affect our analysis in a first 
	approach, because they occur sporadically and randomly and do not 
	affect substantially the periodicities and the power spectra. On the  
	other hand, Figure \ref{fig4} shows significant differences in the period 27 to 30 May in 
	comparison with Figure \ref{fig3}. For instance, this radio source shows 
	appreciable fluctuations on 27 May, when periods of $\approx$ 6 sec ($\approx$ 0.17 Hz) 
	and also $\approx$ 40 sec ($\approx$ 0.025 Hz) are within the cone of influence.  The 
	behavior of the {\it vTEC} and the {\it Dst} Index correspond to a  
	quiet period for this day suggesting that these fluctuations are probably
	associated with ionospheric perturbations restricted to the auroral zone. In fact, on 28  
	May the 
	radio signal shows slight fluctuations with periodicities greater than 
	$\sim$50\,sec\,($\sim$0.02\,Hz), but on 29 May there was an 
	increase in power that concentrated around 20\,sec\,($\sim$0.05\,Hz). 
	This increase could be probably associated with the arrival to Earth 
	of a CME, as is reflected by the {\it Dst} index for these 
	days. Several CMEs took place around this period, but we believe that only the  halo CME  
	observed ($\sim$14:06 UT, with a velocity of 
	427\,km\,s$^{-1}$) on 24 May plus the high-speed stream from a coronal hole 
	observed on the solar disc may be responsible for the geomagnetic storm 
	(see Figure \ref{fig2}). In this scenario, a complete study 
	is always useful to understand the whole context in the interplanetary medium, interplanetary 
	CMEs ({\it e.g.}, ejecta and magnetic cloud), corotating interaction regions or CIRs, etc. Finally, on 
	30 May the behavior of this source radio-signal returns to a quiet state without fluctuations, as 
	shown in Figure \ref{fig4}.  Figure \ref{fig5} shows a different period of observation for this source, from 24 to 28 April. During this period of time, the radio signal  presents fluctuations every day. However, the 
	frequencies indicate different origins. In fact, for 24 and 26 Arpil the {\it vTEC} and the {\it Dst} index display a behavior corresponding to a quiet period (see 
	Figure \ref{fig2}), suggesting that these fluctuations are probably 
	associated with perturbations in the interplanetary medium producing 
	IPS. For 27 and 28 April, the fluctuations could be 
	associated with IPS plus IOND. Figure \ref{fig9}  shows that 
	the index presents an abrupt peak on 30 April. This increase, although due to ionospheric disturbances, as can be appreciated from the wavelet analysis,
does not let us clearly understand what is the contribution of  the ionosphere (in relation with the {\it Dst} variation) because this requires data from more radio sources.
	%However, sometimes ionospheric perturbations can contaminate 
	%the radio-signals as has been shown in the work of \citet{car11}. 
	%These oscillations could be associated with a CME event on April 
	%23 ($\sim23:54$\,hrs UT, CPA$=106^{\circ}$, AW$=57^{\circ}$, and 
	%269\,km\,s$^{-1}$). 
\item {\it 3C405} shows a behavior similar to that of {\it Cas A}, when on 28 May 
	the radio signal presents slight fluctuations with 
	periodicities greater than 50\,sec ($\sim$0.02\,Hz) and with a 
	significant power at these frequencies. However, on 29 May, the 
	fluctuations are stronger with periodicities greater than 
	10\,sec\,($\sim$0.1\,Hz) and also with a significant power at these 
	frequencies (see Figure \ref{fig6}). Because the {\it vTEC} (see 
	Figure \ref{fig10}) is not highly altered locally on 29 May, it is 
	suggested that the fluctuations observed are probably due to 
	ionospheric perturbations around the auroral region. On 30 May the radio signal 
	of this source shows a behavior corresponding to a quiet period. Finally, on 31 May, the radio signal of this source presents slight fluctuations: however,  in this case, the {\it vTEC} (see Figure \ref{fig10}) is also 
	slightly perturbed at the end of 30 May, suggesting  
	that the observed fluctuations may be contaminated by ionospheric 
	disturbances (see Figure \ref{fig6}).

\begin{figure}
\centering%
\includegraphics[width=6.0cm,angle=270]{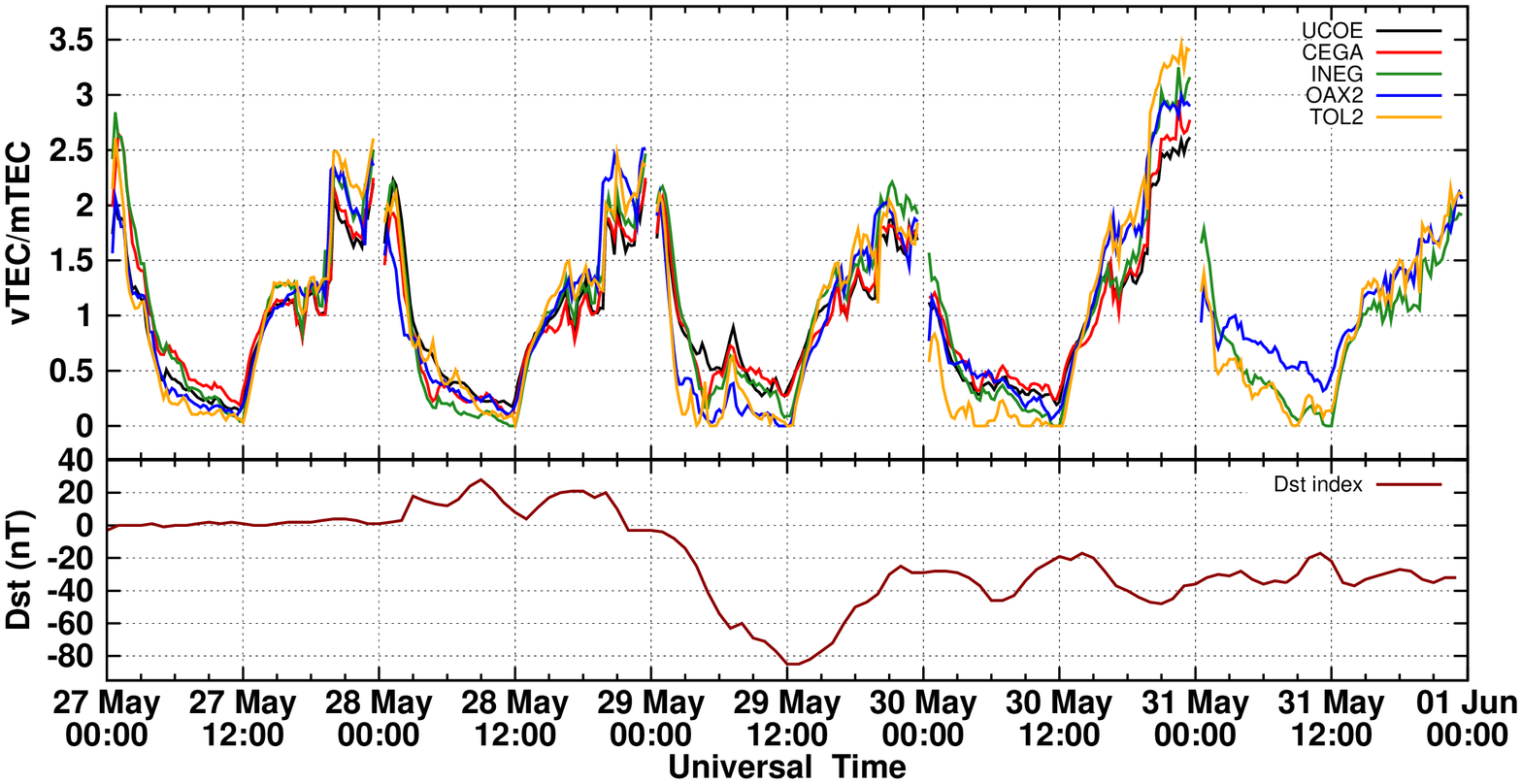}
\caption{The vertical total electron content ({\it vTEC}) and {\it Dst} and the Dst index shown in detail 
from 27 May to 1 June 2010. The top panel shows the {\it vTEC} divided by the mean value of the 
TEC (mTEC) and calculated using data from five stations: UCOE, CEGA, INEG, OAX2 and TOL2, represented with lines in black, red, 
green, blue and orange colors respectively (see section \ref{tec_dstindex}). The bottom panel shows the {\it Dst} 
index.}
\label{fig10}
\end{figure}

\item {\it 3C144}  shows a different behavior. No fluctuations were observed on 27 May (see Figure \ref{fig7}).  However, 
        the radio signal of this source is highly perturbed, the perturbations have periods of  around 
	10\,sec\,($\sim$0.1\,Hz). In addition, both the {\it vTEC}  and the {\it Dst} (see Figure 
	\ref{fig10}) 
	show a behavior typical of quiet periods, suggesting that the fluctuations 
	observed on this day may be due to IPS. There are several CME events 
	between 25 and 28 May that may contribute to the 
	IPS. Finally, on 29 and 30 May no significant 
	fluctuations are observed (see Figure \ref{fig7}).
\item {\it 3C48,} remains 
	unperturbed on 26 and 30 May, but on 28 and 
	29 May there are slight fluctuations observed in the wavelet 
	window. However, the power spectra show no periodicities above
	the 95\% significance level (see Figure 
	\ref{fig8}). In this case, the fluctuations observed could be 
	associated with IPS and ionospheric perturbations, respectively. 
	However, we need a higher S/N to have a better definition of their origin. 
\end{enumerate}

In summary, we have observed several 
sources that present fluctuations in their radio signal (see Table 
\ref{table1}). These fluctuations may be associated with IOND, suggested by the {\it vTEC} in some days; but, in other cases, they were 
associated with IPS, and sometimes with both. The fact that sometimes the 
radio signal from astronomical sources is contamined by ionospheric 
disturbances, suggests that these fluctuations should be taken into account as 
possible sources of contamination in IPS measurements obtained with the MEXART observatory \citep{car11}.

From our results, we have found that this complementary study can help to 
understand the origin of the observed fluctuations in the radio signal of 
astronomical sources. In addition, we have found that the ionospheric 
disturbances can play an important role in the contamination of the signal 
when looking for IPS. The ionosphere acts as an obstacle to radio waves of 
$\leq 150$\,MHz \citep{coh09} and sometimes it is not possible to discriminate 
{\it a priori} between IPS or ionospheric disturbances as sources of the 
observed intensity fluctuations recorded by MEXART. 
 
From a further analysis using the radio signal from satellites transiting 
close to an astronomical source and detected by MEXART observatory, it was possible to 
discern 
whether the intensity fluctuations were associated with IOND or IPS 
\citep{car11}. In this context, we have used an additional tool, a wavelet analysis, in this article. This alternative tool lets us characterize the perturbed signal.

\section{Conclusions}
\label{con}

The conclusions of our work can be summarized as follows:

\begin{itemize}
\item Our study incorporates the use of wavelets together with 
	complementary information provided by the {\it vTEC} and {\it Dst} index. This is the 
	first time these tools have been applied to MEXART data and allow a better 	
	understanding of cases when  the 
	IPS may be contaminated by IOND.
\item Within the period of April--May 2010, several radio sources presented fluctuations in their radio signal registered with the MEXART. Using wavelets we found the periods/frequencies that 
	characterize these fluctuations. We found that the perturbations in the signal were sometimes associated with IOND and/or IPS (see for instance Figure \ref{fig4} top left, Figure \ref{fig5} top left and Figure \ref{fig8} top right). In this context the 
	{\it vTEC} can help to establish the cases in which the IPS is 
	actually contaminated by IOND. Furthermore, in the example 
	shown in Figure \ref{fig8} (top left), the wavelet analysis indicates that the ionospheric 
	scintillation is associated with quiet conditions, and in Figure \ref{fig5} (top left) and Figure 
	\ref{fig6} (top right) the wavelet analysis shows ionospheric disturbances associated with the 
	geomagnetic storm.
\item In addition, Fourier spectra show that these fluctuations have  
	frequencies between $\sim$\,0.01 Hz to $\sim$\,1.0 Hz 
	(periods of 100\,sec to 1\,sec, respectively) for both IOND and 
	IPS. We found that there is a tendency for the frequencies 
	close to $\sim$\,1.0 Hz to be more likely associated with IPS, whereas lower 
	frequencies tend to be related to IOND. However, in the cases  
	where both IPS and IOND occur, we could not distinguish between 
	them.  
\item In the examples that neither the Dst index nor the {\it vTEC} showed any significant change, we 
         can say that the fluctuations are not related to global ionospheric effects but could be 
         probably related to IPS. We observed that when the {\it vTEC} is perturbed and  solar-origin 
         effects can be assumed, then the radio signal can be contaminated by ionospheric 
         perturbations. This is probably the case observed by \citet{car11}, where  
	the fluctuations could be attributed to IOND and/or IPS.
\item The solar-minimum period has allowed us to characterize better the radio fluctuations 
	of astronomical sources observed with the MEXART, during this period the ionosphere can 
	contribute strongly. This is especially true for those sources that having large elongations 
	(around 90$^{\circ}$, for instance) present fluctuations in their signal. However, more work is 
	needed to characterize the intrinsic fluctuations in the radio signal. 
\end{itemize}

\begin{acknowledgements}

The authors thank the MEXART team, in particular:  A. Gonz\'alez-Esparza (P.\,I. of MEXART), P. 
Villanueva, E. Andrade, and E. Aguilar, for the technical support and provision of data. We also 
thank CORS, SOPAC and UNAVCO for incorporating the data on their respective websites from 
which the GPS data can be freely downloaded. We are also grateful to NOAA-SWPC for access to 
the TEC calculation program. We also wish to thank the Servicio Sismol\'ogico Nacional and E. 
Cabral for providing some Mexican GPS data. This article was  possible thanks to the 
PAPIIT-UNAM projects IN111509 and IA102514-2. M. Rodr\'\i guez-Mart\'\i nez also thanks 
REDCyTE and CONACyT for a postdoctoral fellowship supported by the project {\it Estudio de la 
ionosfera y sus aplicaciones} for the Mexican Space Agency (MSA). The authors finally thank the 
suggestions of M. S\'anchez   and S. Kurtz that have improved this article. 

\end{acknowledgements}

\end{document}